\documentclass[aps,onecolumn]{revtex4-2}
\usepackage{graphicx}
\usepackage{epsfig}
\usepackage{epstopdf}
\usepackage{amsfonts}
\usepackage{amssymb}
\usepackage{amsbsy}
\usepackage{amsmath}
\usepackage{mathrsfs}
\usepackage{latexsym}
\usepackage{natbib}
\usepackage{bm}
\usepackage{color}
\usepackage[a4paper, total={6.0in, 9.5in}]{geometry}

\usepackage{braket}
\usepackage{slashed}
\usepackage{pgfplots}
\numberwithin{equation}{section}

\usepackage{tikz}
\usetikzlibrary{shapes,arrows,shadows}

\begin{document}

\title{Thermal field theories coupled to curved spacetime}

\author{Susobhan Mandal}
\email{sm12ms085@gmail.com}

\affiliation{ Department of Physics,\\ 
Indian Institute of Science Education and Research Tirupati,\\
Tirupati -  517507, India }

\date{\today}

\begin{abstract}
Thermal field theory is an essential tool for comprehending various physical phenomena, including astrophysical objects such as neutron stars and white dwarfs, as well as the early stages of the universe. Nonetheless, the traditional thermal field theory formulated in Minkowski spacetime is not capable of considering the effects originating from the curved spacetime. These effects are crucial for both astrophysical and cosmological observations, making it essential to extend the domain of thermal field theory to curved spacetimes. This article's primary focus is to explore the extension of thermal field theory to curved spacetimes and its implications. We employ Riemann-normal coordinates to describe thermal field theories in curved spacetime, and we also calculate several thermodynamic observables to demonstrate the curvature corrections explicitly.  
\end{abstract}

\maketitle

\section{Introduction}

Quantum field theory is an established framework that describes the behavior of many-particle systems at zero temperature. Despite its accuracy in matching experimental observations in high-energy physics, it has limitations when applied to real-world phenomena, which occur at non-zero temperatures. To enhance the accuracy of theoretical predictions, it is crucial to consider the contributions of temperature. This is where thermal field theory \cite{laine2016basics, altherr1993introduction, Kapusta:2006pm} becomes invaluable. Thermal field theory enables us to understand the temperature-dependent corrections to physical observables, playing a vital role in the evolution of important systems such as white dwarfs, neutron stars, and early stages of the universe. In such systems, the temperature is sufficiently high that any theoretical predictions require explicit use of thermal field theory. Thermal field theory becomes successful as it helps us to comprehend the phenomena of phase transitions and cosmological inflation in the early universe, the evolution of a neutron star, and other compact objects.

%
%
%
Thermal field theory has been successful in accounting for temperature-dependent corrections to physical observables and predicting various phenomena, such as phase transitions. However, it is only applicable when the background spacetime geometry is Minkowski flat. Consequently, the reliability of its predictions for systems ranging from the early universe to compact astrophysical objects \cite{tsuruta1979thermal, potekhin2001thermal, blandford1983thermal, potekhin2003thermal, perez2006anisotropic, ivanova2004thermal, townsley2004theoretical}, including neutron stars, white dwarfs, and boson stars, is questionable since it fails to consider corrections due to the background curved spacetime. For these extremely dense objects, it is crucial to consider curvature-dependent and other background geometry corrections to accurately describe their physics. In the case of early universe physics and cosmology \cite{kolb1981early, schleicher2008reionization, giudice1999thermal, ellis1996thermal, fornengo1997finite, farrar1993baryon, vilenkin1982gravitational, hogan1983nucleation}, the expansion of the universe is a significant factor in many phenomena. Therefore, accurately predicting theoretical outcomes requires accounting for the effects of the scale factor of the universe.
%

In order to consider the corrections coming from the background curved spacetime, we use the formulation of QFT in curved spacetime using Riemann-normal coordinates (RNC), shown in \cite{mandal2019local, bunch1979feynman}. In \cite{mandal2019local}, it is shown explicitly that computation techniques in the RNC formulation mimic the computation techniques used in conventional QFT in Minkowski spacetime. The current theme of this article is to extend the RNC formalism of QFT in curved spacetime such that it is also applicable in thermal field theories assuming local thermal equilibrium such that one can use the imaginary time formulation. We also discuss some of its consequences in terms of physical observables namely, curvature corrections to the heat capacity and pressure exerted by scalar bosons and fermions in free-field theories from the free-energy, Bose-Einstein condensation phenomenon, curvature induced conserved 4-current associated with global $U(1)$ symmetry in complex scalar field theory and finally the curvature correction to Coleman-Weinberg Potential. 

\section{Conventional thermal field theory}
\subsection{Scalar field theory}
In thermal field theory, the partition function is an important quantity from which most of the observables can be derived, as it plays the role of a generating functional. Partition function 
of a field theory in equilibrium with temperature $(k_{B}\beta)^{-1}$ is defined as
\begin{equation}
\mathcal{Z}=\text{Tr}[e^{-\beta\hat{H}}],
\end{equation}
where $\hat{H}$ is the Hamiltonian of the system and using the functional integral formalism, 
this can be expressed as
\begin{equation}
\mathcal{Z}=\int_{\phi(\tau,\vec{x})=\phi(\tau+\beta,\vec{x})}\mathcal{D}\phi \ e^{-S_{E}[\phi]},
\end{equation}
where $S_{E}[\phi]$ is the euclidean action. $S_{E}[\phi]$ (upto a minus sign) can be derived from the canonical action by doing a Wick rotation $t\rightarrow-i\tau$. As a consequence of the periodic boundary condition $\phi(\tau,\vec{x})=\phi(\tau+\beta,\vec{x})$, the scalar field can be expressed as
\begin{equation}\label{Fourier1}
\phi(\tau,\vec{x})=\sqrt{\frac{\beta}{V}}\sum_{n}\sum_{\vec{p}}\phi(\omega_{n},\vec{p})e^{i(\omega_{n}\tau+\vec{p}.\vec{x})},
\end{equation}
with $\omega_{n}=\frac{2\pi n}{\beta}$, known as Matsubara frequencies. For a free-scalar 
field theory, the euclidean action can be expressed as
\begin{equation}\label{action 1}
S_{E}[\phi]=\int_{0}^{\beta}d\tau\int d^{3}x\Big[\frac{1}{2}(\partial_{\tau}\phi)^{2}+\frac{1}{2}(\vec{\nabla}\phi)^{2}+\frac{1}{2}m^{2}\phi^{2}\Big].
\end{equation}  
Substituting the representation in (\ref{Fourier1}) into (\ref{action 1}), the partition function can be expressed as
\begin{equation}\label{partition function1}
\begin{split}
\mathcal{Z} & =\int\mathcal{D}\phi \ e^{-\frac{\beta^{2}}{2}\sum_{n,\vec{p}}(\omega_{n}^{2}+\vec{p}^{2}+m^{2})\phi(-\omega_{n},-\vec{p})\phi(\omega_{n},\vec{p})}=\mathcal{N}\prod_{n,\vec{p}}(\beta^{2}(\omega_{n}^{2}+\vec{p}^{2}+m^{2}))^{-\frac{1}{2}},
\end{split}
\end{equation}
where $\mathcal{N}$ is an overall normalization factor which is irrelevant for practical purposes. Hence, we can write $\log\mathcal{Z}$ in following form
\begin{equation}
\begin{split}
\log\mathcal{Z} & =-\frac{1}{2}\sum_{n,\vec{p}}\log(\beta^{2}(\omega_{n}^{2}+\omega_{\vec{p}}^{2})), \ \omega_{\vec{p}}=\sqrt{\vec{p}^{2}+m^{2}}\\
 & =-\frac{1}{2}\sum_{\vec{p}}\int_{1}^{\beta\omega_{\vec{p}}}dx\left(\frac{1}{2}+\frac{2}{e^{x}-1}\right)=V\int\frac{d^{3}p}{(2\pi)^{3}}\Big[-\frac{1}{2}\beta\omega_{\vec{p}}-\log(1-e^{-\beta\omega_{\vec{p}}})\Big],
\end{split}
\end{equation}
where the first term in \textit{r.h.s} is the zero-point energy in zero-temperature (divergent in nature) which needs to be subtracted to get thermal contribution to $\log\mathcal{Z}$. From $\log\mathcal{Z}$, we can derive pressure as follows
\begin{equation}
p=\frac{\beta^{-1}}{V}\log\mathcal{Z}=-\frac{1}{\beta}\int\frac{d^{3}p}{(2\pi)^{3}}\log(1-e^{-\beta\omega_{\vec{p}}}),
\end{equation}
which is reduced to following simple form for massless field theories
\begin{equation}
p=\frac{\pi^{2}}{90\beta^{4}}.
\end{equation} 
Note that the expression for partition function in (\ref{partition function1}) can also be expressed as
\begin{equation}\label{result1}
\begin{split}
\mathcal{Z} & =\mathcal{N}[\text{det}(-\beta^{2}(\Box-m^{2}))]^{-\frac{1}{2}}=\mathcal{N}[\text{det}(\beta^{-2}\mathcal{G}(x))]^{\frac{1}{2}}=\mathcal{N}\prod_{n,\vec{p}}[\beta^{-2}\mathcal{G}(\omega_{n},\vec{p})]^{\frac{1}{2}},
\end{split}
\end{equation}
where $\mathcal{G}(x)$ is the Green's function of scalar field \textit{w.r.t} the periodic boundary condition. Expression (\ref{result1}) is important for deriving $\log\mathcal{Z}$ in case of thermal field theory in a generic curved spacetime.

\subsection{Fermionic field theory}
For free-fermionic field theory, the euclidean action is given by
\begin{equation}
S_{E}[\bar{\psi},\psi]=\int_{0}^{\beta}d\tau\int d^{3}x[\bar{\psi}(\gamma^{0}\partial_{\tau}-i\gamma^{i}\partial_{i}-m)\psi].
\end{equation} 
Partition function for the fermionic field theory can be expressed as a functional integral as follows
\begin{equation}
\mathcal{Z}=\int\mathcal{D}\bar{\psi}\mathcal{D}\psi \ e^{-S_{E}[\bar{\psi},\psi]},
\end{equation}
with anti-periodic boundary conditions on the fermionic field variables $\psi(\tau+\beta,\vec{x})=-\psi(\tau,\vec{x}), \ \bar{\psi}(\tau+\beta,\vec{x})=-\bar{\psi}(\tau,\vec{x})$. Hence, the fermionic fields can also be expressed as
\begin{equation}
\psi(\tau,x)=\sqrt{\frac{1}{V}}\sum_{n,\vec{k}}\psi(\omega_{n},\vec{k})e^{i(\omega_{n}\tau+\vec{k}.\vec{x})}, \ \bar{\psi}(\tau,x)=\sqrt{\frac{1}{V}}\sum_{n,\vec{k}}\bar{\psi}(\omega_{n},\vec{k})e^{-i(\omega_{n}\tau+\vec{k}.\vec{x})},
\end{equation}
with fermionic Matsubara frequencies $\omega_{n}=\frac{(2n+1)\pi}{\beta}$. Using the above representation, the action can be expressed as
\begin{equation}
S_{E}[\bar{\psi},\psi]=\sum_{n,\vec{p}}\bar{\psi}(\omega_{n},\vec{p})(i\gamma^{0}\omega_{n}+\vec{\gamma}.\vec{p}+m)\psi(\omega_{n},\vec{p}).
\end{equation}
Using the above form of the euclidean action, partition function can be expressed as
\begin{equation}
\mathcal{Z}=\prod_{n,\vec{p}}(\omega_{n}^{2}+\omega_{\vec{p}}^{2})^{2}.
\end{equation}
Free-fermionic fields theory is invariant under global $U(1)$ transformation. Hence, in order to maintain the conservation of $U(1)$ charge, chemical potential $\mu$ is introduced in the definition of partition function $\mathcal{Z}=\text{Tr}e^{-\beta(\hat{H}-\mu\hat{N})}$ where $\hat{N}$ is the number operator which is conserved $U(1)$ charge. In the presence of chemical potential, the expression for partition function gets modified as follows
\begin{equation}
\mathcal{Z}=\prod_{n,\vec{p}}[\beta^{2}(\tilde{\omega}_{n}^{2}+\omega_{\vec{p}}^{2})]^{2}, \ \tilde{\omega}_{n}=\omega_{n}+i\mu.
\end{equation} 
From the above expression of $\mathcal{Z}$, we obtain
\begin{equation}\label{free fermions}
\begin{split}
\log\mathcal{Z} & =2V\sum_{n}\int\frac{d^{3}p}{(2\pi)^{3}}\log[\beta^{2}(\tilde{\omega}_{n}^{2}+\omega_{\vec{p}}^{2})]=2\int\frac{d^{3}p}{(2\pi)^{3}}[\beta\omega_{\vec{p}}+\log(1+e^{-\beta(\omega_{\vec{p}}-\mu)})+\log(1+e^{-\beta(\omega_{\vec{p}}+\mu)})]
\end{split}
\end{equation}  
which for massless field theory in degeneracy limit ($\beta\mu\gg1$), becomes following
\begin{equation}
\log\mathcal{Z}=\frac{\beta V}{24\pi^{2}}\Big[2\mu^{4}+\frac{48\mu^{2}}{\beta^{2}}\Big]+\frac{7\pi^{2}}{180\beta^{3}}V
\end{equation}
Thus, the expression for pressure exerted by massless degenerate fermions is
\begin{equation}
P=\frac{1}{24\pi^{2}}\Big[2\mu^{4}+\frac{48\mu^{2}}{\beta^{2}}\Big]+\frac{7\pi^{2}}{180\beta^{4}}
\end{equation}

\section{Thermal field theory in curved spacetimes}
\subsection{Green's function in RNC formalism}
Riemann-normal coordinates \cite{hatzinikitas2000note, petrov2016einstein, Parker:1983pe, Bunch:1981tr}, defined in curved spacetime are the closest analogue of flat spacetime coordinates. This coordinate system is constructed in such a way that geodesics starting from one point to another point of the manifold are mapped to the tangent space. In general RNC covers a patch in the neighborhood of a given point, which will be taken as the origin of this coordinate system and this coordinate system is well-defined as long as the geodesics do not intersect in that patch. A generic point $x$ in RNC patch is defined by the components of the tangent vector (evaluated at the origin of that coordinate system) to the geodesic which links $x$ and $x'$, where $x'$ is chosen to be the origin of this coordinate frame. Denoting the components of the tangent vector by $z^{\mu}$ and $s$ the affine parameter of the geodesic measured between $x$ and $x'$, Riemann-normal coordinates are defined as follows
\begin{equation}
\begin{split}
x^{\mu} & =sz^{\mu}, \ z^{\mu}=-\nabla_{x}^{\mu}\sigma(x,x'),
\end{split}
\end{equation}  
where $\sigma(x,x')$ is Synge's world function \cite{vines2015geodesic, rylov1990extremal}.

Using RNC, the definition of Green's function containing the d'Alembertian operator can be 
expanded in following way
\begin{equation}
(-g(x))^{\frac{1}{4}}(-\Box_{x}+m^{2}+\xi\mathcal{R}(x))\mathcal{G}(x,y)(-g(y))^{\frac{1}{4}}=-\delta^{(4)}(x-y).
\end{equation}
Let $\bar{\mathcal{G}}(x,y)\equiv(-g(y))^{\frac{1}{4}}\mathcal{G}(x,y)(-g(x))^{\frac{1}
{4}}$ then using RNC, above equation for Feynman propagator becomes \cite{bunch1979feynman}
\begin{equation} \label{eqn.6}
\begin{split}
\eta^{\mu\nu}\partial_{\mu}\partial_{\nu}\bar{\mathcal{G}} & -\Big[m^{2}+(\xi-\frac{1}{6})\Big]\bar{\mathcal{G}}-\frac{1}{3}\mathcal{R}_{\alpha}^{ \ \nu}z^{\alpha}\partial_{\nu}\bar{\mathcal{G}}+\frac{1}{3}\mathcal{R}_{ \ \alpha \ \beta}^{\mu \ \nu}z^{\alpha}z^{\beta}\partial_{\mu}\partial_{\nu}\bar{\mathcal{G}}\\
-(\xi-\frac{1}{6})\mathcal{R}_{;\alpha}z^{\alpha}\bar{\mathcal{G}} & +(\frac{1}{6}\mathcal{R}_{\alpha\beta;}^{ \ \ \nu}-\frac{1}{3}\mathcal{R}_{\alpha \ \ ;\beta}^{ \ \nu})z^{\alpha}z^{\beta}\partial_{\nu}\bar{\mathcal{G}}+\frac{1}{6}\mathcal{R}_{ \ \alpha \ \beta;\gamma}^{\mu \ \nu}z^{\alpha}z^{\beta}z^{\gamma}\partial_{\mu}\partial_{\nu}\bar{\mathcal{G}}\\
-\frac{1}{2}(\xi-\frac{1}{6})\mathcal{R}_{;\alpha\beta}z^{\alpha}z^{\beta}\bar{\mathcal{G}} & +(\frac{1}{40}\Box \mathcal{R}_{\alpha\beta}-\frac{1}{120}\mathcal{R}_{;\alpha\beta}-\frac{1}{130}\mathcal{R}_{\alpha}^{ \ \mu}\mathcal{R}_{\beta\mu}-\frac{1}{60}\mathcal{R}^{\lambda\kappa}\mathcal{R}_{\lambda\alpha\beta\kappa}+\frac{1}{60}\mathcal{R}_{ \ \alpha}^{\lambda \ \mu\kappa}\mathcal{R}_{\lambda\beta\mu\kappa})z^{\alpha}z^{\beta}\bar{\mathcal{G}}\\
+(\frac{1}{10}\mathcal{R}_{\alpha\beta; \ \gamma}^{ \ \ \nu}-\frac{3}{20} & \mathcal{R}_{\alpha \ ;\beta\gamma}^{ \ \nu}+\frac{1}{60}\mathcal{R}_{ \ \alpha}^{\lambda}\mathcal{R}_{\lambda\beta\gamma}^{ \ \ \ \nu}-\frac{1}{15}\mathcal{R}_{ \ \alpha\beta}^{\lambda \ \ \kappa}\mathcal{R}_{\lambda\gamma \ \kappa}^{ \ \ \nu})z^{\alpha}z^{\beta}z^{\gamma}\partial_{\nu}\bar{\mathcal{G}}\\
+(\frac{1}{20}\mathcal{R}_{ \ \alpha \ \beta;\gamma\delta}^{\mu \ \nu} & +\frac{1}{15}\mathcal{R}_{ \ \alpha\beta}^{\lambda \ \ \mu}R_{\lambda\gamma\delta}^{ \ \ \ \nu})z^{\alpha}z^{\beta}z^{\gamma}z^{\delta}\partial_{\mu}\partial_{\nu}\bar{\mathcal{G}}+\ldots=-\delta^{(4)}(x-y).
\end{split}
\end{equation}
On the other Green's function can also be expressed as
\begin{equation}
\bar{\mathcal{G}}(z)=\int\frac{d^{4}k}{(2\pi)^{4}}e^{ik_{\mu}z^{\mu}}\bar{\mathcal{G}}(k).
\end{equation}
In the momentum space, (\ref{eqn.6}) can be solved by taking into account derivatives of the metric in ascending order as
\begin{equation}\label{green function expansion}
\bar{\mathcal{G}}(k)=\bar{\mathcal{G}}_{0}(k)+\bar{\mathcal{G}}_{1}(k)+\bar{\mathcal{G}}_{2}(k)+\ldots,
\end{equation}
where $\bar{\mathcal{G}}_{i}(k)$ contains $i^{\text{th}}$ derivative of metric at $z=0$ and is of the order of $k^{-(2+i)}$ such that dimensional wise all the terms match.

The first few terms of the above expansion are following 
\begin{equation}\label{curved green function}
\begin{split}
\bar{\mathcal{G}}_{0}(k) & =\frac{1}{k^{2}+m^{2}}, \ \bar{\mathcal{G}}_{1}(k)=0, \ \bar{\mathcal{G}}_{2}(k)=-\frac{(\xi-\frac{1}{6})\mathcal{R}}{(k^{2}+m^{2})^{2}}, \ \bar{\mathcal{G}}_{3}(k)=i\frac{(\xi-\frac{1}{6})\mathcal{R}_{;\alpha}}{k^{2}+m^{2}}\partial^{\alpha}(k^{2}+m^{2})^{-1}\\
\bar{\mathcal{G}}_{4}(k) & =(\xi-\frac{1}{6})^{2}\mathcal{R}^{2}(k^{2}+m^{2})^{-3}+\frac{1}{2}(\xi-\frac{1}{6})\frac{\mathcal{R}_{;\alpha\beta}}{k^{2}+m^{2}}\partial^{\alpha}\partial^{\beta}(k^{2}+m^{2})^{-1}\\
 & -a_{\alpha\beta}(k^{2}+m^{2})^{-1}\partial^{\alpha}\partial^{\beta}(k^{2}+m^{2})^{-1},
\end{split}
\end{equation}
where
\begin{equation}
a_{\alpha\beta}=\frac{1}{40}\Box \mathcal{R}_{\alpha\beta}-\frac{1}{120}\mathcal{R}_{;\alpha\beta}-\frac{1}{30}\mathcal{R}_{\alpha}^{ \ \mu}\mathcal{R}_{\beta\mu}-\frac{1}{60}\mathcal{R}^{\lambda\kappa}\mathcal{R}_{\lambda\alpha\beta\kappa}+\frac{1}{60}\mathcal{R}_{ \ \alpha}^{\lambda \ \mu\kappa}\mathcal{R}_{\lambda\beta\mu\kappa}.
\end{equation}
Here all the geometrical quantities (Riemann tensor, Ricci tensor, Ricci scalar and their various derivatives) are evaluated at origin $z=0$. The expression for Green's function in a generic curved spacetime, derived in (\ref{green function expansion}), (\ref{curved green function}) using RNC formalism is important for deriving the expression for $\log\mathcal{Z}$ in that curved spacetime. 

\subsection{Thermal scalar field theory in curved spacetime}
Plugging the expansion of Green's function (\ref{green function expansion}) in (\ref{result1}), we obtain the following expression
\begin{equation}
\begin{split}
\mathcal{Z} & =\mathcal{N}\prod_{n,\vec{p}}[\beta^{-2}\mathcal{G}(\omega_{n},\vec{p})]^{\frac{1}{2}}=\mathcal{N}\prod_{n,\vec{p}}[\beta^{-2}(\mathcal{G}_{0}(\omega_{n},\vec{p})+\mathcal{G}_{2}(\omega_{n},\vec{p})+\mathcal{G}_{3}(\omega_{n},\vec{p})+\ldots)]^{\frac{1}{2}}\\
 & =\mathcal{N}\prod_{n,\vec{p}}[\beta^{-2}\mathcal{G}_{0}(\omega_{n},\vec{p})(1+\mathcal{G}_{0}^{-1}(\omega_{n},\vec{p})\mathcal{G}_{2}(\omega_{n},\vec{p})+\mathcal{G}_{0}^{-1}(\omega_{n},\vec{p})\mathcal{G}_{3}(\omega_{n},\vec{p})+\ldots)]^{\frac{1}{2}}.
\end{split}
\end{equation}
Thus, the expression for $\log\mathcal{Z}$ takes the following form
\begin{equation}
\begin{split}
\log\mathcal{Z} & =(\log\mathcal{Z})_{0}+(\log\mathcal{Z})_{1}\\
(\log\mathcal{Z})_{0} & =-\frac{1}{2}\sum_{n,\vec{p}}\log(\beta^{2}(\omega_{n}^{2}+\omega_{\vec{p}}^{2}))\\
(\log\mathcal{Z})_{1} & =\frac{1}{2}\sum_{n,\vec{p}}\log(1+\mathcal{G}_{0}^{-1}(\omega_{n},\vec{p})\mathcal{G}_{2}(\omega_{n},\vec{p})+\mathcal{G}_{0}^{-1}(\omega_{n},\vec{p})\mathcal{G}_{3}(\omega_{n},\vec{p})+\ldots),
\end{split}
\end{equation}
and thus pressure can be expressed as
\begin{equation}
P=P_{0}+P_{1}, \ \ P_{0}=\frac{\beta^{-1}}{V}(\log\mathcal{Z})_{0}, \ P_{1}=\frac{\beta^{-1}}{V}(\log\mathcal{Z})_{1}.
\end{equation}
For massless field theory $P_{0}=\frac{\pi^{2}}{90\beta^{4}}$ which is the result coming from pure Minkwoski geometry, shown earlier. On the other hand, $P_{1}$ is the contribution to the pressure due to background curved spacetime. Denoting $\tilde{\mathcal{G}}_{i}(\omega_{n},\vec{p})\equiv\mathcal{G}_{0}^{-1}(\omega_{n},\vec{p})\mathcal{G}_{i}(\omega_{n},\vec{p})$, we compute $P_{1}$ perturbatively
\begin{equation}\label{expansion for P1}
\begin{split}
P_{1} & =\frac{\beta^{-1}}{2V}\sum_{n,\vec{p}}\log(1+\mathcal{G}_{0}^{-1}(\omega_{n},\vec{p})\mathcal{G}_{2}(\omega_{n},\vec{p})+\mathcal{G}_{0}^{-1}(\omega_{n},\vec{p})\mathcal{G}_{3}(\omega_{n},\vec{p})+\ldots)\\
 & =\frac{\beta^{-1}}{2}\sum_{n}\int\frac{d^{3}p}{(2\pi)^{3}}\Big[\tilde{\mathcal{G}}_{2}(\omega_{n},\vec{p})+\tilde{\mathcal{G}}_{3}(\omega_{n},\vec{p})+\tilde{\mathcal{G}}_{4}(\omega_{n},\vec{p})-\frac{1}{2}\tilde{\mathcal{G}}_{2}^{2}(\omega_{n},\vec{p})+\ldots\Big],
\end{split}
\end{equation} 
of which $\sum_{n}\int\frac{d^{3}p}{(2\pi)^{3}}\tilde{\mathcal{G}}_{3}(\omega_{n},\vec{p})=0$ since it is odd in $k^{i},\omega_{n}$. Hence, we need to compute first, third and fourth terms in the series expansion for $P_{1}$. The first term in (\ref{expansion for P1}) is expressed as
\begin{equation}\label{P11}
\begin{split}
P_{1}^{(1)} & \equiv\frac{\beta^{-1}}{2}\sum_{n}\int\frac{d^{3}p}{(2\pi)^{3}}\tilde{\mathcal{G}}_{2}(\omega_{n},\vec{p})=-\frac{\beta^{-1}}{2}\left(\xi-\frac{1}{6}\right)\mathcal{R}\sum_{n}\int\frac{d^{3}p}{(2\pi)^{3}}\frac{1}{\omega_{n}^{2}+\omega_{\vec{p}}^{2}}\\
 & =-\frac{\beta^{-1}}{2}\left(\xi-\frac{1}{6}\right)\mathcal{R}\beta\int_{\mathcal{C}}\frac{dp_{0}}{2\pi i}\int\frac{d^{3}p}{(2\pi)^{3}}\frac{1}{(p_{0})^{2}-\omega_{\vec{p}}^{2}}\coth\left(\frac{\beta p_{0}}{2}\right)\\
 & =-\frac{\beta^{-1}}{2}\left(\xi-\frac{1}{6}\right)\mathcal{R}\beta\int\frac{d^{3}p}{(2\pi)^{3}}\frac{1}{2\omega_{\vec{p}}}\left(1+\frac{2}{e^{\beta\omega_{\vec{p}}}-1}\right)\\
\implies P_{1, \ \text{phys}}^{(1)} & =-\frac{\beta^{-1}}{2}\left(\xi-\frac{1}{6}\right)\mathcal{R}\beta\int\frac{d^{3}p}{(2\pi)^{3}}\frac{1}{\omega_{\vec{p}}}\frac{1}{e^{\beta\omega_{\vec{p}}}-1}=-\frac{\beta^{-1}}{2}\left(\xi-\frac{1}{6}\right)\mathcal{R}\frac{1}{2\pi^{2}\beta}\int_{0}^{\infty} dx \ \frac{x}{e^{x}-1}\\
 & =-\left(\xi-\frac{1}{6}\right)\frac{\mathcal{R}}{4\pi^{2}\beta^{2}}\zeta(2)\implies P_{1, \ \text{phys}}^{(1)}|_{\xi=0}=\frac{\mathcal{R}}{24\pi^{2}\beta^{2}}\zeta(2),
\end{split}
\end{equation}
where $P_{1, \ \text{phys}}^{(1)}$ is the non-divergent piece. On the other hand, the fourth term in (\ref{expansion for P1}) is expressed as
\begin{equation}\label{P14}
\begin{split}
P_{1}^{(4)} & \equiv -\frac{\beta^{-1}}{4}\sum_{n}\int\frac{d^{3}p}{(2\pi)^{3}}\tilde{\mathcal{G}}_{2}^{2}(\omega_{n},\vec{p})=-\frac{\beta^{-1}}{4}\left(\xi-\frac{1}{6}\right)^{2}\mathcal{R}^{2}\sum_{n}\int\frac{d^{3}p}{(2\pi)^{3}}\left(\frac{1}{\omega_{n}^{2}+\omega_{\vec{p}}^{2}}\right)^{2}\\
 & =\frac{1}{4}\left(\xi-\frac{1}{6}\right)^{2}\mathcal{R}^{2}\int\frac{d^{3}p}{(2\pi)^{3}}\frac{\partial}{\partial p^{2}}\int_{\mathcal{C}}\frac{dp_{0}}{2\pi i}\Big[\frac{1}{(p_{0})^{2}-\omega_{\vec{p}}^{2}}\coth\left(\frac{\beta p_{0}}{2}\right)\Big]\\
 & =\frac{1}{4}\left(\xi-\frac{1}{6}\right)^{2}\mathcal{R}^{2}\int\frac{d^{3}p}{(2\pi)^{3}}\frac{\partial}{\partial p^{2}}\Big[\frac{1}{2\omega_{\vec{p}}}\left(1+\frac{2}{e^{\beta\omega_{\vec{p}}}-1}\right)\Big]\\
P_{1, \ \text{phys}}^{(4)} & \equiv\frac{1}{4}\left(\xi-\frac{1}{6}\right)^{2}\mathcal{R}^{2}\int\frac{d^{3}p}{(2\pi)^{3}}\frac{\partial}{\partial p^{2}}\Big[\frac{1}{p}\frac{1}{e^{\beta p}-1}\Big]=\frac{1}{16\pi^{2}}\left(\xi-\frac{1}{6}\right)^{2}\mathcal{R}^{2}\int_{0}^{\infty}p \ dp \frac{\partial}{\partial p}\Big[\frac{1}{p}\frac{1}{e^{\beta p}-1}\Big].
\end{split}
\end{equation}
The zero-temperature contribution in pressure is divergent, hence, those contributions are removed in the physical finite-temperature contribution. A similar divergent contribution is obtained for the first term in $\frac{\beta^{-1}}{2}\sum_{n}\int\frac{d^{3}p}{(2\pi)^{3}}\tilde{\mathcal{G}}_{4}(\omega_{n},\vec{p})$ hence, it is also unphysical. On the other hand, the second term in $\tilde{\mathcal{G}}_{4}(\omega_{n},\vec{p})$ would not contribute to the pressure in massless field theory which follows from the following identities
\begin{equation}\label{P13}
\begin{split}
(k^{2}+m^{2})^{-1} & \partial^{\alpha}\partial^{\beta}(k^{2}+m^{2})^{-1}=\frac{1}{k^{2}+m^{2}}\Big[-\frac{2\delta^{\alpha\beta}}{(k^{2}+m^{2})^{2}}+\frac{8k^{\alpha}k^{\beta}}{(k^{2}+m^{2})^{3}}\Big]\\
\int\frac{d^{4}k}{(2\pi)^{4}} & \frac{1}{k^{2}+m^{2}}\Big[-\frac{2\delta^{\alpha\beta}}{(k^{2}+m^{2})^{2}}+\frac{8k^{\alpha}k^{\beta}}{(k^{2}+m^{2})^{3}}\Big]=-2m^{2}\delta^{\alpha\beta}\int\frac{d^{4}k}{(2\pi)^{4}}\frac{1}{(k^{2}+m^{2})^{4}}.
\end{split}
\end{equation}
Thus from (\ref{P11}), (\ref{P14}) and (\ref{P13}), we obtain following expression for pressure upto second order curvature corrections
\begin{equation}\label{total pressure}
P=\frac{\pi^{2}}{90\beta^{4}}+\frac{\mathcal{R}}{144\beta^{2}}+\mathcal{O}(\beta^{2}).
\end{equation}
From the expression (\ref{total pressure}), we obtain the following expression for specific heat capacity of the system
\begin{equation}\label{specific heat capacity1}
\begin{split}
C_{V} & =\frac{\partial}{\partial T}\left(-\frac{1}{V}\frac{\partial}{\partial\beta}\log\mathcal{Z}\right)=\frac{\partial}{\partial T}\Big[\frac{\pi^{2}}{30\beta^{4}}+\frac{\mathcal{R}}{144\beta^{2}}\Big]=k_{B}\left(\frac{4\pi^{2}}{30\beta^{3}}+\frac{\mathcal{R}}{72\beta}\right).
\end{split}
\end{equation}
Hence, we derive the expressions for pressure and specific heat capacity in (\ref{total pressure}) and (\ref{specific heat capacity1}) respectively for a non-interacting thermal scalar field theory in curved spacetime. From the expression for specific heat capacity in (\ref{specific heat capacity1}), it follows that heat capacity at a given point $x$, becomes negative i.e $C_{V}(x)<0$ if $\beta^{2}(x)<-\frac{288\pi^{2}}{30\mathcal{R}(x)}$ provided $\mathcal{R}(x)<0$. 

\section{Thermal field theory for fermions in curved spacetime}
\subsection{Tetrad formalism in curved spacetime}
Given a generic curved spacetime manifold with metric $g_{\mu\nu}(x)$, local 1-form basis can be written in terms of vierbein in following way
\begin{equation}
\begin{split}
\omega^{a}(x) & =e_{\mu}^{ \ a}(x)dx^{\mu}\\
\implies g_{\mu\nu}(x) & =e_{\mu}^{ \ a}(x)e_{\nu}^{ \ b}(x)\eta_{ab}.
\end{split}
\end{equation} 
These vierbeins are invertible which implies there exists $e_{ \ a}^{\mu}(x)$ such that
\begin{equation}
e_{ \ a}^{\mu}(x)e_{\nu}^{ \ a}(x)=\delta_{\nu}^{\mu}, \ e_{ \ a}^{\mu}(x)e_{\mu}^{ \ b}(x)=\delta_{a}^{b}.
\end{equation}
A covariant derivative can be defined for these local field variables in terms of spin connection defined as follows
\begin{equation}
\nabla_{\mu}A^{a}(x)=\partial_{\mu}A^{a}(x)+\omega_{\mu \ \ b}^{ \ a}(x)A^{b}(x).
\end{equation}
This connection must be compatible with vierbein i.e. it must satisfy the condition $\nabla_{\mu}e_{\nu}^{ \ a}(x) = 0$. This together with Leibniz rule fix the form of the spin-connection in terms of Christoffel symbols as follows
\begin{equation}\label{spin-connection}
\omega_{\mu \ \ b}^{ \ a}=-e_{ \ b}^{\nu}(\partial_{\mu}e_{\nu}^{ \ a}-\Gamma_{ \ \mu\nu}^{\lambda}e_{\lambda}^{ \ a}).
\end{equation}
In terms of (\ref{spin-connection}), we define the following one-form
\begin{equation}\label{1-form}
\omega_{ \ b}^{a}=\omega_{\mu \ \ b}^{ \ a}dx^{\mu},
\end{equation}
The exterior derivative of the one-form $\omega^{a}$ is expressed as
\begin{equation}
\begin{split}
d\omega^{a} & =e_{[\nu,\mu]}^{ \ \ a}dx^{\mu}\wedge dx^{\nu}\\
e_{[\nu,\mu]}^{ \ \ a} & =\frac{1}{2}(\partial_{\mu}e_{\nu}^{ \ a}-\partial_{\nu}e_{\mu}^{ \ a}).
\end{split}
\end{equation}
From the Cartan's equation for curvature form, expressed as
\begin{equation}
d\omega^{a}+\omega_{ \ b}^{a}\wedge\omega^{b}=0,
\end{equation}
and the torsionless condition of Christoffel symbols imply that
\begin{equation}
\omega_{\mu ba}=-\omega_{\mu ab}.
\end{equation}
Using the one-form in (\ref{1-form}), the curvature two-form can be expressed using Cartan' approach and we obtain
\begin{equation}
\begin{split}
\mathcal{R}_{ \ b}^{a} & =d\omega_{ \ b}^{a}+\omega_{ \ c}^{a}\wedge\omega_{ \ b}^{c}\equiv-\frac{1}{2}\mathcal{R}_{\mu\nu \ b}^{ \ \ a}dx^{\mu}\wedge dx^{\nu}\\
\mathcal{R}_{\mu\nu \ b}^{ \ \ a} & =-[\partial_{\mu}\omega_{\nu \ b}^{ \ a}-\partial_{\nu}\omega_{\mu \ b}^{ \ a}+\omega_{\mu \ c}^{a}\omega_{\nu \ b}^{c}-\omega_{\nu \ c}^{a}\omega_{\mu \ b}^{c}].
\end{split}
\end{equation}
Using the spin-connection, the covariant derivative for spinors is defined as
\begin{equation}
\nabla_{\mu}\Psi=\partial_{\mu}\Psi+i\omega_{\mu}^{ \ ab}\Sigma_{ab}\Psi, \ \Sigma^{ab}=-\frac{i}{8}[\gamma^{a},\gamma^{b}].
\end{equation} 
From the $SO(1,3)$ algebra
\begin{equation}
[\Sigma_{ab},\Sigma_{cd}]=\frac{i}{2}(\eta_{ac}\Sigma_{bd}-\eta_{ad}\Sigma_{bc}-\eta_{bc}\Sigma_{ad}+\eta_{bd}\Sigma_{ac}),
\end{equation}
it follows that
\begin{equation}
[\nabla_{\mu},\nabla_{\nu}]\Psi=-i\mathcal{R}_{\mu\nu}^{ \ \ ab}\Sigma_{ab}\Psi=-\frac{1}{8}\mathcal{R}_{\mu\nu}^{ \ \ ab}[\gamma_{a},\gamma_{b}]\Psi.
\end{equation}
The above relation has an important consequence. Using this result together with following identities
\begin{equation}
\begin{split}
\gamma^{\mu}\gamma^{\nu} & (\nabla_{\mu}\nabla_{\nu}-\nabla_{\nu}\nabla_{\mu})\Psi=-\frac{1}{4}\gamma^{\mu}\gamma^{\nu}\gamma^{\lambda}\gamma^{\sigma}\mathcal{R}_{\mu\nu\lambda\sigma}\Psi\\
\mathcal{R}_{\mu\nu\lambda\sigma} & +\mathcal{R}_{\mu\lambda\nu\sigma}+\mathcal{R}_{\mu\sigma\nu\lambda}=0,
\end{split}
\end{equation}
it can be easily shown that
\begin{equation}
(\gamma^{\mu}\nabla_{\mu})^{2}=\left(-\Box+\frac{1}{4}\mathcal{R}\right)\mathbb{I}_{4\times4}, \ \Box=\frac{1}{\sqrt{-g}}\partial_{\mu}(\sqrt{-g}g^{\mu\nu}\partial_{\nu}).
\end{equation}

\subsection{Dirac field theory in curved spacetime}
The Dirac action in a generic curved spacetime is given by
\begin{equation}
S=\int d^{4}x\sqrt{-g}\Big[\frac{i}{2}(\bar{\Psi}\gamma^{\mu}\nabla_{\mu}\Psi-\overline{\nabla_{\mu}\Psi}\gamma^{\mu}\Psi)-m\bar{\Psi}\Psi\Big].
\end{equation}
Then the partition function is expressed as
\begin{equation}
\mathcal{Z}=\int\mathcal{D}\bar{\Psi}\mathcal{D}\Psi \ e^{iS}=\text{det}(l\mathfrak{D}), \ \mathfrak{D}=(i\gamma^{\mu}\nabla_{\mu}-m),
\end{equation}
and free-energy is expressed as
\begin{equation}
\log\mathcal{Z}=\log\text{det}(l\mathfrak{D}).
\end{equation}
The arbitrary length scale $l$ is introduced to make the argument of $\text{det}(.)$ dimensionless and all the physical observable is independent of $l$. In order to determine the expression for $\log\mathcal{Z}$, we define following operator
\begin{equation}
\tilde{\mathfrak{D}}=-i\gamma^{\mu}\nabla_{\mu}-m.
\end{equation}
Recall that $\gamma_{5}\equiv i\gamma^{0}\gamma^{1}\gamma^{2}\gamma^{3}$ satisfy following identities
\begin{equation}
\{\gamma_{\mu},\gamma_{5}\}=0,\ \gamma_{5}^{2}=\mathbb{I},
\end{equation}
and hence 
\begin{equation}\label{result}
\tilde{\mathfrak{D}}=\gamma_{5}\mathfrak{D}\gamma_{5}, \ \text{det}(l\mathfrak{D})=\text{det}(l\tilde{\mathfrak{D}}).
\end{equation}
Using (\ref{result}), we can express $\log\mathcal{Z}$ as follows
\begin{equation}\label{free fermions curved}
\begin{split}
\log\mathcal{Z} & =\frac{1}{2}\log[\text{det}(l\mathfrak{D})\text{det}(l\mathfrak{D})]=\frac{1}{2}\log[\text{det}(l\mathfrak{D})\text{det}(l\tilde{\mathfrak{D}})]\\
 & =\frac{1}{2}\log[\text{det}(l^{2}\mathfrak{D}\tilde{\mathfrak{D}})]=2\log[\text{det}[l^{2}(-\Box+m^{2}+\frac{1}{4}\mathcal{R})]],
\end{split} 
\end{equation}
where we have used the fact that $\mathfrak{D}\tilde{\mathfrak{D}}=\tilde{\mathfrak{D}}\mathfrak{D}=\Big[(\gamma^{\mu}\nabla_{\mu})^{2}+m^{2}\Big]\mathbb{I}_{4\times4}=\left(-\Box+m^{2}+\frac{1}{4}\mathcal{R}\right)\mathbb{I}_{4\times4}$.  

\subsection{Thermal fermionic field theory in curved spacetime}
In the case of fermionic field theory, eigenvalues of operator $\left(-\Box+m^{2}+\frac{1}{4}\mathcal{R}\right)$ is the same as in the case of scalar field theory but with fermionic Matsubara frequencies. From the results in (\ref{free fermions}) and (\ref{free fermions curved}), it follows that
\begin{equation}
\begin{split}
\log\mathcal{Z} & =(\log\mathcal{Z})_{0}+(\log\mathcal{Z})_{1}\\
(\log\mathcal{Z})_{0} & =2\sum_{n,\vec{p}}\log(\beta^{2}(\omega_{n}^{2}+\omega_{\vec{p}}^{2}))\\
(\log\mathcal{Z})_{1} & =-2\sum_{n,\vec{p}}\log(1+\mathcal{G}_{F, 0}^{-1}(\omega_{n},\vec{p})\mathcal{G}_{F, 2}(\omega_{n},\vec{p})+\mathcal{G}_{F, 0}^{-1}(\omega_{n},\vec{p})\mathcal{G}_{F, 3}(\omega_{n},\vec{p})+\ldots),
\end{split}
\end{equation}
where expression of $(\log\mathcal{Z})_{0}$ is already given in (\ref{free fermions}). Hence, pressure is given by
\begin{equation}
P=P_{0}+P_{1}, \ \ P_{0}=\frac{\beta^{-1}}{V}(\log\mathcal{Z})_{0}, \ P_{1}=\frac{\beta^{-1}}{V}(\log\mathcal{Z})_{1}.
\end{equation} 
Denoting $\tilde{\mathcal{G}}_{F, i}(\omega_{n},\vec{p})\equiv\mathcal{G}_{F, 0}^{-1}(\omega_{n},\vec{p})\mathcal{G}_{F, i}(\omega_{n},\vec{p})$, we compute $P_{1}$ perturbatively
\begin{equation}\label{expansion for P1 fermion}
\begin{split}
P_{1} & =-\frac{2\beta^{-1}}{V}\sum_{n,\vec{p}}\log(1+\mathcal{G}_{F, 0}^{-1}(\omega_{n},\vec{p})\mathcal{G}_{F, 2}(\omega_{n},\vec{p})+\mathcal{G}_{F,0}^{-1}(\omega_{n},\vec{p})\mathcal{G}_{F, 3}(\omega_{n},\vec{p})+\ldots)\\
 & =-2\beta^{-1}\sum_{n}\int\frac{d^{3}p}{(2\pi)^{3}}\Big[\tilde{\mathcal{G}}_{F, 2}(\omega_{n},\vec{p})+\tilde{\mathcal{G}}_{F, 4}(\omega_{n},\vec{p})-\frac{1}{2}\tilde{\mathcal{G}}_{F, 2}^{2}(\omega_{n},\vec{p})+\ldots\Big],
\end{split}
\end{equation}
where we used the fact that $\sum_{n}\int\frac{d^{3}p}{(2\pi)^{3}}\tilde{\mathcal{G}}_{F, 3}(\omega_{n},\vec{p})=0$. Now we compute the first few terms in the series expansion in (\ref{expansion for P1 fermion}). The first term in the series is expressed as
\begin{equation}
\begin{split}
P_{1}^{(1)} & \equiv- 2\beta^{-1}\sum_{n}\int\frac{d^{3}p}{(2\pi)^{3}}\tilde{\mathcal{G}}_{2}(\omega_{n},\vec{p})=2\beta^{-1}\left(\xi-\frac{1}{6}\right)\mathcal{R}\sum_{n}\int\frac{d^{3}p}{(2\pi)^{3}}\frac{1}{\omega_{n}^{2}+\omega_{\vec{p}}^{2}}\\
 & =2\beta^{-1}\left(\xi-\frac{1}{6}\right)\mathcal{R}\beta\int_{\mathcal{C}}\frac{dp_{0}}{2\pi i}\int\frac{d^{3}p}{(2\pi)^{3}}\frac{1}{(p_{0})^{2}-\omega_{\vec{p}}^{2}}\tanh\left(\frac{\beta(p_{0}-\mu)}{2}\right)\\
 & =2\beta^{-1}\left(\xi-\frac{1}{6}\right)\mathcal{R}\beta\int\frac{d^{3}p}{(2\pi)^{3}}\frac{1}{2\omega_{\vec{p}}}\left(1-\frac{2}{e^{\beta(\omega_{\vec{p}}-\mu)}+1}\right)\\
\implies P_{1, \ \text{phys}}^{(1)} & =-2\left(\xi-\frac{1}{6}\right)\mathcal{R}\int\frac{d^{3}p}{(2\pi)^{3}}\frac{1}{\omega_{\vec{p}}}\frac{1}{e^{\beta(\omega_{\vec{p}}-\mu)}+1}\\
 & =-\frac{\left(\xi-\frac{1}{6}\right)}{\pi^{2}}\mathcal{R}\int_{0}^{\mu}p \ dp-\frac{\left(\xi-\frac{1}{6}\right)}{\pi^{2}}\mathcal{R}\int_{0}^{\infty}dp\frac{p}{e^{\beta p}+1}, \ \text{for} \ (\beta\mu\gg1)\\
 & =-\frac{\left(\xi-\frac{1}{6}\right)}{2\pi^{2}}\mathcal{R}\mu^{2}-\frac{\left(\xi-\frac{1}{6}\right)}{\pi^{2}\beta^{2}}\mathcal{R}\eta(2), \ \eta(2)=\sum_{n=1}^{\infty}(-1)^{n-1}\frac{1}{n^{2}}\\
\implies P_{1, \ \text{phys}}^{(1)}|_{\xi=\frac{1}{4}} & =-\frac{\mu^{2}}{24\pi^{2}}\mathcal{R}-\frac{1}{288\beta^{2}}\mathcal{R}, 
\end{split}
\end{equation}
where $P_{1, \ \text{phys}}^{(1)}$ is the non-divergent piece. On the other hand, the second and third terms together are expressed as
\begin{equation}
\begin{split}
P_{1}^{(2+3)} & \equiv -\beta^{-1}\sum_{n}\int\frac{d^{3}p}{(2\pi)^{3}}\tilde{\mathcal{G}}_{2}^{2}(\omega_{n},\vec{p})=-\beta^{-1}\left(\xi-\frac{1}{6}\right)^{2}\mathcal{R}^{2}\sum_{n}\int\frac{d^{3}p}{(2\pi)^{3}}\left(\frac{1}{\omega_{n}^{2}+\omega_{\vec{p}}^{2}}\right)^{2}\\
 & =\left(\xi-\frac{1}{6}\right)^{2}\mathcal{R}^{2}\int\frac{d^{3}p}{(2\pi)^{3}}\frac{\partial}{\partial p^{2}}\Big[\frac{1}{2\omega_{\vec{p}}}\left(1-\frac{2}{e^{\beta(\omega_{\vec{p}}-\mu)}+1}\right)\Big]\\
P_{1}^{(2+3)'} & \equiv -\left(\xi-\frac{1}{6}\right)^{2}\mathcal{R}^{2}\int\frac{d^{3}p}{(2\pi)^{3}}\frac{\partial}{\partial p^{2}}\Big[\frac{1}{p}\frac{1}{e^{\beta(p-\mu)}+1}\Big]=-\frac{1}{4\pi^{2}}\left(\xi-\frac{1}{6}\right)^{2}\mathcal{R}^{2}\int_{0}^{\infty}p \ dp \frac{\partial}{\partial p}\Big[\frac{1}{p}\frac{1}{e^{\beta(p-\mu)}+1}\Big]\\
 & =\frac{1}{4\pi^{2}}\left(\xi-\frac{1}{6}\right)^{2}\mathcal{R}^{2}\Bigg[\frac{1}{e^{-\beta\mu}+1}-\underbrace{\int_{0}^{\infty}\frac{dp}{p}\frac{1}{e^{\beta(p-\mu)}-1}}_{divergent}\Bigg]\\
\implies P_{1, \ \text{phys}}^{(2+3)} & \equiv\frac{1}{4\pi^{2}}\left(\xi-\frac{1}{6}\right)^{2}\mathcal{R}^{2}\frac{1}{e^{-\beta\mu}+1}. 
\end{split}
\end{equation}
In the expression of $P_{1, \ \text{phys}}^{(2+3)}$, all the divergent terms are removed. Thus in fermionic degeneracy limit, the net pressure upto second order curvature correction is expressed as
\begin{equation}
\begin{split}
P & = \frac{1}{24\pi^{2}}\Big[2\mu^{4}+\frac{48\mu^{2}}{\beta^{2}}\Big]+\frac{7\pi^{2}}{180\beta^{4}} 
- \left(\xi - \frac{1}{6}\right)\frac{\mu^{2}}{2\pi^{2}}\mathcal{R}\\
 & - \left(\xi - \frac{1}{6}\right)\frac{\eta(2)}{\pi^{2}\beta^{2}}\mathcal{R} + \frac{1}{4\pi^{2}}\left(\xi-\frac{1}{6}\right)^{2}\mathcal{R}^{2}\frac{1}{e^{-\beta\mu}+1},
\end{split}
\end{equation}
whereas the specific heat capacity is given by
\begin{equation}
\begin{split}
C_{V} & =\frac{\partial}{\partial T}\left(-\frac{1}{V}\frac{\partial}{\partial\beta}\log\mathcal{Z}\right)\\
 & =\frac{\partial}{\partial T}\Big[\frac{7\pi^{2}}{60\beta^{4}}+\frac{2\mu^{2}}{\pi^{2}\beta^{2}}-\frac{1}{288\beta^{2}}\mathcal{R}+\frac{1}{576\pi^{2}}\mathcal{R}^{2}\frac{e^{-\beta\mu}(\beta\mu-1)-1}{(e^{-\beta\mu}+1)^{2}}\Big]\\
 & =k_{B}\left(\frac{7\pi^{4}}{15\beta^{3}}+\frac{4\mu^{2}}{\pi^{2}\beta}-\frac{1}{144\beta}\mathcal{R}\right)+\mathcal{O}(\mathcal{R}^{2}).
\end{split}
\end{equation}

%

\section{Bose-Einstein condensation in curved spacetime}
Below a critical temperature, particles in a Bose gas (follows Bose-Einstein distribution) can occupy the same quantum ground state, forming a Bose-Einstein condensate (BEC). BEC is an important phenomenon that takes place in compact astrophysical objects \cite{chavanis2012bose, latifah2014bosons, potekhin2003thermal}, and under weak two-body interaction, this generates a superfluid state of matter in these compact objects.
\subsection{BEC in non-interacting charged scalar field in Minkowski spacetime}
Global $U(1)$-invariant action for a non-interacting charged scalar field theory is given by
\begin{equation}
S[\bar{\phi},\phi]=\int d^{4}x \ (\partial_{\mu}\bar{\phi}\partial^{\mu}\phi-m^{2}\bar{\phi}\phi).
\end{equation}
Since, it is invariant under the action of global $U(1)$-group, there exists a conserved current given by
\begin{equation}
j^{\mu}=i(\bar{\phi}\partial^{\mu}\phi-\partial^{\mu}\bar{\phi}\phi).
\end{equation}
The partition function for this field theory is given by following functional integral
\begin{equation}\label{partition BEC}
\mathcal{Z} = \int\mathcal{D}\bar{\pi}\mathcal{D}\pi\int\mathcal{D}\bar{\phi}\mathcal{D}\phi  \ e^{\int_{0}^{\beta}d\tau\int d^{3}x(\bar{\pi}\partial_{\tau}\phi+\pi\partial_{\tau}\bar{\phi}-\mathcal{H}+\mu j^{0})},
\end{equation}
where $\mathcal{H}$ is the hamiltonian density. In terms of the following parametrization
\begin{equation}
\phi=\frac{1}{\sqrt{2}}(\phi_{1}+i\phi_{2}), \ \pi=\frac{1}{\sqrt{2}}(\pi_{1}+i\pi_{2}),
\end{equation} 
the exponent in the functional integral in (\ref{partition BEC}) becomes
\begin{equation}
\begin{split}
\bar{\pi}\partial_{\tau}\phi & +\pi\partial_{\tau}\bar{\phi}-\mathcal{H}+\mu j^{0}=\frac{1}{2}[-\{(\partial_{\tau}\phi_{1})^{2}+\partial_{\tau}\phi_{2})^{2}+(\vec{\nabla}\phi_{1})^{2}+(\vec{\nabla}\phi_{2})^{2}+(m^{2}-\mu^{2})(\phi_{1}^{2}+\phi_{2}^{2})\}\\
 & +2i\mu(\phi_{2}\partial_{\tau}\phi_{1}-\phi_{1}\partial_{\tau}\phi_{2})]-\frac{1}{2}[(\pi_{1}-i\partial_{\tau}\phi_{1}-\mu\phi_{2})^{2}+(\pi_{2}-i\partial_{\tau}\phi_{2}+\mu\phi_{1})^{2}]\\
 & \equiv -\mathcal{L}_{E}-\frac{1}{2}[\tilde{\pi}_{1}^{2}+\tilde{\pi}_{2}^{2}],
\end{split}
\end{equation}
where $\tilde{\pi}_{1}=\pi_{1}-i\partial_{\tau}\phi_{1}-\mu\phi_{2}$ and $\tilde{\pi}_{2}=\pi_{2}-i\partial_{\tau}\phi_{2}+\mu\phi_{1}$.

Using the momentum space representation of fields
\begin{equation}
\phi_{i}(x)=\zeta_{i}+\sqrt{\frac{\beta}{V}}\sum_{n,\vec{k}}e^{-i(\omega_{n}\tau+\vec{k}.\vec{x})}\phi_{i}(\omega_{n},\vec{k}),
\end{equation}
the euclidean action takes following form
\begin{equation}
\begin{split}
S_{E}[\phi_{1},\phi_{2}] & =\beta V\frac{m^{2}-\mu^{2}}{2}\zeta^{2}+\frac{1}{2}\sum_{n,\vec{k}}\begin{bmatrix}
\phi_{1}(-\omega_{n},-\vec{k}) & \phi_{2}(-\omega_{n},-\vec{k})
\end{bmatrix}\beta^{2}\mathbf{D}_{0}(\omega_{n},\vec{k})\begin{bmatrix}
\phi_{1}(\omega_{n},\vec{k})\\
\phi_{2}(\omega_{n},\vec{k})
\end{bmatrix}\\
\mathbf{D}_{0}(\omega_{n},\vec{k}) & =\begin{bmatrix}
\omega_{n}^{2}+\omega_{\vec{k}}^{2}-\mu^{2} & 2\mu\omega_{n}\\
-2\mu\omega_{n} & \omega_{n}^{2}+\omega_{\vec{k}}^{2}-\mu^{2}
\end{bmatrix},
\end{split}
\end{equation}
where $\zeta^{2}=\zeta_{1}^{2}+\zeta_{2}^{2}$, and $\zeta_{i}$ is the zero momentum mode of the field $\phi_{i}$. After doing the functional integral over $\pi_{1}, \ \pi_{2}$ fields, partition function becomes (up to an irrelevant overall multiplicative factor)
\begin{equation}\label{charged scalar partition}
\begin{split}
\mathcal{Z} & =\int\mathcal{D}\phi_{1}\mathcal{D}\phi_{2} \ e^{-S_{E}[\phi_{1},\phi_{2}]}\\
\implies\log\mathcal{Z} & =\beta V\frac{\mu^{2}-m^{2}}{2}\zeta^{2}-\frac{1}{2}\sum_{n,\vec{k}}\log[\beta^{4}((\omega_{n}^{2}+\omega_{\vec{k}}^{2}-\mu^{2})^{2}+4\mu^{2}\omega_{n}^{2})]\\
 & =\beta V\frac{\mu^{2}-m^{2}}{2}\zeta^{2}-\frac{1}{2}\sum_{n,\vec{k}}\log[\beta^{4}((\omega_{n}+i\mu)^{2}+\omega_{\vec{k}}^{2})((\omega_{n}-i\mu)^{2}+\omega_{\vec{k}}^{2})]\\
 & =\beta V\frac{\mu^{2}-m^{2}}{2}\zeta^{2}-V\int\frac{d^{3}k}{(2\pi)^{3}}\Big[\beta\omega_{\vec{k}}+\log(1-e^{-\beta(\omega_{\vec{k}}-\mu)})+\log(1-e^{-\beta(\omega_{\vec{k}}+\mu)})\Big],
\end{split}
\end{equation}
where the second term in the parenthesis in the last line refers to the contribution coming from the particles whereas the third term refers to the contribution coming from the anti-particles. In order to avoid complex values of free-energy, $|\mu|\leq m$ must hold. The renormalized thermodynamic free-energy density is given by
\begin{equation}
\Omega=\frac{m^{2}-\mu^{2}}{2}\zeta^{2}+\int\frac{d^{3}k}{(2\pi)^{3}}\Big[\log(1-e^{-\beta(\omega_{\vec{k}}-\mu)})+\log(1-e^{-\beta(\omega_{\vec{k}}+\mu)})\Big].
\end{equation}
Further, the charge densities are given by
\begin{equation}\label{charge densities}
\begin{split}
Q_{+} & =\int\frac{d^{3}k}{(2\pi)^{3}}\frac{1}{e^{\beta(\omega_{\vec{k}}-\mu)}-1}, \ Q_{-}=\int\frac{d^{3}k}{(2\pi)^{3}}\frac{1}{e^{\beta(\omega_{\vec{k}}+\mu)}-1},
\end{split}
\end{equation}
and from the above expressions for charge densities, it can be shown that
\begin{equation}
Q_{+}|_{\beta\rightarrow0}=Q_{-}|_{\beta\rightarrow0}\approx\frac{\zeta(3)}{\pi^{2}\beta^{3}}, \ (Q_{+}-Q_{-})_{\beta\rightarrow\infty}\approx\left(\frac{m}{2\pi\beta}\right)^{\frac{3}{2}}e^{\beta\mu},
\end{equation}
which shows that in small temperature there is a huge difference in the population of particles and anti-particles because of $\beta\mu\gg1$ whereas for large temperature population difference becomes zero $(Q_{+}-Q_{-})_{\beta\rightarrow 0}=0$. The chemical potential favors particles to be more populated at low temperature if $\mu>0$ and favors anti-particles to be more populated at low temperature if $\mu<0$. 

From the expression of $\Omega$, it follows that for low temperature if $\mu=m$ then an arbitrary number of particles can occupy the ground state without increasing free-energy at all, which is known as Bose-Einstein condensation. This happens when ($\mu\rightarrow m$) the density of system reaches a critical density. Using minimization principle of free-energy \textit{w.r.t} $\zeta$, it can be easily shown that for $m>|\mu|$, $\zeta=0$ whereas for $m=\mu$, $\zeta$ is undetermined which in general takes a non-zero value once interaction is introduced in the system. The value of $\zeta$ is given by
\begin{equation}
\zeta^{2}=\frac{1}{\mu}\left(Q-\int\frac{d^{3}k}{(2\pi)^{3}}\frac{1}{e^{\beta(\omega_{\vec{k}}-\mu)}-1}+\int\frac{d^{3}k}{(2\pi)^{3}}\frac{1}{e^{\beta(\omega_{\vec{k}}+\mu)}-1}\right),
\end{equation}
which follows from $Q=Q_{+}+Q_{-}=-\frac{\partial\Omega}{\partial\mu}$ and critical temperature $T_{c}$ is defined by imposing $\zeta=0$ at $\mu=m$
\begin{equation}
Q=\int\frac{d^{3}k}{(2\pi)^{3}}\frac{1}{e^{\beta_{c}(\omega_{\vec{k}}-m)}-1}-\int\frac{d^{3}k}{(2\pi)^{3}}\frac{1}{e^{\beta_{c}(\omega_{\vec{k}}+m)}-1},
\end{equation}
which can be simplified using a Taylor series expansion
\begin{equation}
\begin{split}
Q & =-2m\int\frac{d^{3}k}{(2\pi)^{3}}\frac{\partial}{\partial\omega_{\vec{k}}}\frac{1}{e^{\beta_{c}\omega_{\vec{k}}}-1}\Big|_{m=0}+\mathcal{O}((\beta_{c}^{-2}m)^{3})\\
 & \approx\frac{\beta_{c}m}{\pi^{2}}\int_{0}^{\infty} dk \frac{k^{2}e^{\beta_{c}k}}{(e^{\beta_{c}k}-1)^{2}}=\frac{m}{\pi^{2}\beta_{c}^{2}}\int_{0}^{\infty}dx\frac{x^{2}e^{x}}{(e^{x}-1)^{2}}=\frac{m}{3\beta_{c}^{2}}\\
\implies & \beta_{c}^{-1}=\sqrt{\frac{3Q}{m}}.
\end{split}
\end{equation}
Thus, if $Q\ll1$ then $T_{c}\ll1$ in which case above approximation becomes exact.

\subsection{BEC in non-interacting charged scalar field coupled to curved spacetime}
For a non-interacting charged scalar field, coupled to a generic curved spacetime, expression for $\log\mathcal{Z}$ in (\ref{charged scalar partition}) can be extended to following form
\begin{equation}
\begin{split}
\log\mathcal{Z} & =\beta V\frac{\mu^{2}-m^{2}}{2}\zeta^{2}+\frac{1}{2}\sum_{n,\vec{k}}\log[\beta^{2}\mathcal{G}(\omega_{n,+},\vec{k})]+\frac{1}{2}\sum_{n,\vec{k}}\log[\beta^{2}\mathcal{G}(\omega_{n,-},\vec{k})]\\
 & \equiv(\log\mathcal{Z})_{0}+(\log\mathcal{Z})_{1}\\
(\log\mathcal{Z})_{1} & =\frac{V}{2}\sum_{n}\int\frac{d^{3}p}{(2\pi)^{3}}[\tilde{\mathcal{G}}_{2}(\omega_{n,+},\vec{k})+\tilde{\mathcal{G}}_{2}(\omega_{n,-},\vec{k})+\ldots],
\end{split}
\end{equation}
where $\omega_{n,\pm}=\omega_{n}\pm i\mu, \ \omega_{n}=\frac{2n\pi}{\beta}$. Using the result in (\ref{expansion for P1}), renormalized $(\log\mathcal{Z})_{1}$ upto leading order curvature correction, is expressed as
\begin{equation}
\begin{split}
(\log\mathcal{Z})_{1} & =\frac{V}{12}\mathcal{R}\beta\int\frac{d^{3}p}{(2\pi)^{3}}\frac{1}{\omega_{\vec{p}}}\left(\frac{1}{e^{\beta(\omega_{\vec{p}}-\mu)}-1}+\frac{1}{e^{\beta(\omega_{\vec{p}}+\mu)}-1}\right)\\
 & =\frac{V}{24\pi^{2}}\mathcal{R}\beta\int_{0}^{\infty}dp \ p \Bigg[\left(\frac{1}{e^{\beta(p-\mu)}-1}+\frac{1}{e^{\beta(p+\mu)}-1}\right)+\mathcal{O}(m\beta)\Bigg]\\
 & =\frac{V}{24\pi^{2}}\mathcal{R}\beta\Big[\frac{\pi^{2}}{3\beta^{2}}-\frac{\mu^{2}}{2}\Big]+\ldots.
\end{split}
\end{equation}
Thus the renormalized expression for $\Omega$ upto leading order curvature correction becomes
\begin{equation}
\begin{split}
\Omega & =\frac{m^{2}-\mu^{2}}{2}\zeta^{2}+\int\frac{d^{3}k}{(2\pi)^{3}}\Big[\log(1-e^{-\beta(\omega_{\vec{k}}-\mu)})+\log(1-e^{-\beta(\omega_{\vec{k}}+\mu)})\Big]-\frac{1}{24\pi^{2}}\mathcal{R}\Big[\frac{\pi^{2}}{3\beta^{2}}-\frac{\mu^{2}}{2}\Big],
\end{split}
\end{equation}
and hence, $Q$ becomes
\begin{equation}
\begin{split}
Q & =-\frac{\partial\Omega}{\partial\mu}=\zeta^{2}\mu+\int\frac{d^{3}k}{(2\pi)^{3}}\frac{1}{e^{\beta(\omega_{\vec{k}}-\mu)}-1}-\int\frac{d^{3}k}{(2\pi)^{3}}\frac{1}{e^{\beta(\omega_{\vec{k}}+\mu)}-1}-\frac{\mu}{24\pi^{2}}\mathcal{R}.
\end{split}
\end{equation} 
Again, the critical temperature $T_{c}$ can be found out by imposing the condition that $\zeta(T_{c})=0$ at $\mu=m$ which implies
\begin{equation}
\begin{split}
Q & =\int\frac{d^{3}k}{(2\pi)^{3}}\frac{1}{e^{\beta(\omega_{\vec{k}}-\mu)}-1}-\int\frac{d^{3}k}{(2\pi)^{3}}\frac{1}{e^{\beta(\omega_{\vec{k}}+\mu)}-1}-\frac{m}{24\pi^{2}}\mathcal{R}\\
 & =\frac{m}{3\beta_{c}^{2}}-\frac{m}{24\pi^{2}}\mathcal{R}\\
\implies\beta_{c}^{-1} & =\sqrt{\frac{3Q}{m}+\frac{\mathcal{R}}{8\pi^{2}}}. 
\end{split}
\end{equation}
The above expression is the leading order curvature correction to the critical temperature below which BEC condensation happens.  Further, since $T_{c}\geq0$ which imposes the constraint $\frac{3Q}{m}+\frac{\mathcal{R}}{8\pi^{2}}\geq0$.

\subsection{U(1) current in curved spacetime}
Here, we show that expectation value of conserved 4-current associated to $U(1)$ symmetry is non-zero at zero-temperature in a generic curved spacetime. We also discuss why that is the case. For simplicity, we consider free-complex scalar field theory which is invariant under global $U(1)$ transformation. The associated conserved 4-current is given by
\begin{equation}
J_{\mu}(x)=\frac{i}{2}(\bar{\phi}(x)\partial_{\mu}\phi(x)-\partial_{\mu}\bar{\phi}(x)\phi(x)),
\end{equation} 
which in momentum space can be expressed as
\begin{equation}
J_{\mu}(x)=\int\frac{d^{4}k}{(2\pi)^{4}}k_{\mu}\bar{\phi}(k)\phi(k).
\end{equation}
Thus, the expectation value of $J_{\mu}(k)$ is given by
\begin{equation}
\begin{split}
\langle J_{\mu}(x)\rangle & =\int\frac{d^{4}k}{(2\pi)^{4}}k_{\mu}\langle\bar{\phi}(k)\phi(k)\rangle=\frac{i}{6}\mathcal{R}_{;\alpha}\int\frac{d^{4}k}{(2\pi)^{4}}\frac{k_{\mu}k^{\alpha}}{(k^{2}+m^{2})^{3}}+\ldots\\
 & =-\frac{1}{24}\mathcal{R}_{;\mu}\int\frac{d^{4}k}{(2\pi)^{4}}\frac{k^{2}}{(k^{2}+m^{2})^{3}}+\ldots=\left(\frac{1}{384\pi^{2}}\mathcal{R}_{;\mu}+\ldots\right)+\text{divergent terms},
\end{split}
\end{equation}
where $\ldots$ represents higher-order curvature correction. After adding the suitable counterterms, we obtain vacuum expectation value of $U(1)$ current in the leading order to be
\begin{equation}\label{non-zero current}
\langle J_{\mu}(x)\rangle=\frac{1}{384\pi^{2}}\mathcal{R}_{;\mu}.
\end{equation} 
The above expression shows that in order to have a non-zero vacuum expectation value of $U(1)$ current without any interaction, $\mathcal{R}_{;\alpha}$ must be non-zero. In a spherical symmetric spacetime $\mathcal{R}$ only depends on the radial component which implies $\langle J_{r}(x)\rangle\neq0$ but $\langle J_{\theta,\varphi}(x)\rangle=0$ provided $\mathcal{R}$ is not a constant. However, for a rotating star where spherical symmetry is broken due to rotation $\langle J_{\theta,\varphi}(x)\rangle\neq0$. Given this leading-order expression of $\langle J_{\mu}(x)\rangle$, we can in principle solve Maxwell's equations
\begin{equation}
\nabla_{\mu}F^{\mu\nu}(x)=\langle J^{\nu}(x)\rangle=\frac{1}{384\pi^{2}}\mathcal{R}^{;\nu}(x),
\end{equation}
which give the distribution of electric and magnetic field in the astrophysical compact objects \cite{gupta1998rotating, rezzolla2001stationary}. This shows a mechanism which could generate magnetic and electric field from the Ricci scalar of the background geometry. 

\section{Coleman-Weinberg mechanism in curved spacetime}
Spontaneous symmetry breaking \cite{Beekman_2019} is a phenomenon, occurs in many-body systems in which the ground state does not respect the symmetry of many-body hamiltonian. Superfluidity and BEC are also closely associated with spontaneous symmetry breaking. In \cite{coleman1973radiative, malbouisson1996new}, it is shown that radiative corrections in massless $\phi^{4}$-theory leads to the occurrence of spontaneous symmetry breaking (SSB). In this section, we discuss the modifications (if there is any) to the Coleman-Weinberg mechanism in curved spacetime due to radiative corrections. We may note that the corrections to effective potential due to curved spacetime has been computed in literature \cite{Hu:1984js, Sobreira:2011ep, dosReis:2018xxr, Gorbar:2003yp} following some other technique.
\subsection{Effective action}
Generating functional of a field theory is expressed as
\begin{equation}\label{generating0}
\mathcal{Z}[J]=\int\mathcal{D}\phi \ e^{iS[\phi]+i\int\sqrt{-g(x)}d^{4}x \ J(x)\phi(x)},
\end{equation}
from which any $n$-point function can be computed by taking functional derivatives \textit{w.r.t} field variables
\begin{equation}\label{Green's function}
\mathcal{G}(x_{1},\ldots,x_{n})=\frac{1}{i^{n}}\frac{1}{\sqrt{-g(x_{1})}\ldots\sqrt{-g(x_{n})}}\frac{\delta^{n}\mathcal{Z}[J]}{\delta J(x_{1})\ldots\delta J(x_{n})}\Big|_{J=0}.
\end{equation}
Using the above expression, generating functional can be expressed as a series expansion in sources $J(x)$
\begin{equation}
\mathcal{Z}[J]=\sum_{n=0}^{\infty}\frac{i^{n}}{n!}\int\sqrt{-g(x_{1})}\ldots\sqrt{-g(x_{n})}\mathcal{G}(x_{1},\ldots,x_{n}) J(x_{1})\ldots J(x_{n}) \ d^{4}x_{1}\ldots d^{4}x_{n}.
\end{equation}
From the generating functional $\mathcal{Z}[J]$, generating functional $\mathcal{W}[J]$ for connected correlation functional is obtained from
\begin{equation}\label{generating2}
\mathcal{Z}[J]=e^{i\mathcal{W}[J]},
\end{equation}
with the connected correlations are expressed as
\begin{equation}
\mathcal{G}_{c}(x_{1},\ldots,x_{n})=\frac{1}{i^{n-1}}\frac{1}{\sqrt{-g(x_{1})}\ldots\sqrt{-g(x_{n})}}\frac{\delta^{n}\mathcal{W}[J]}{\delta J(x_{1})\ldots\delta J(x_{n})}\Big|_{J=0}.
\end{equation}
From (\ref{generating2}), classical field is defined as
\begin{equation}\label{classical field}
\Phi(x)\equiv\langle\phi(x)\rangle_{J(x)}=\frac{\delta\mathcal{W}[J]}{\delta J(x)},
\end{equation}
and solving (\ref{classical field}), $J(x)$ can be expressed as a function of $\Phi(x)$. Expressing $J(x)$ in terms of $\Phi(x)$, we can do a Legendre transformation in order to define effective action
\begin{equation}\label{effective action}
\Gamma[\Phi]=\mathcal{W}[J]-\int\sqrt{-g(x)}J(x)\Phi(x) \ d^{4}x,
\end{equation}
from which, classical field can also be expressed
\begin{equation}
J(x)=-\frac{1}{\sqrt{-g(x)}}\frac{\delta\Gamma[\Phi]}{\delta\Phi(x)}.
\end{equation} 
Effective action in (\ref{effective action}) can also be expressed as a series expansion
\begin{equation}\label{series effective action}
\begin{split}
\Gamma[\Phi] & =\sum_{n=0}^{\infty}\frac{1}{n!}\int\sqrt{-g(x_{1})}\ldots\sqrt{-g(x_{n})}\Gamma^{(n)}(x_{1},\ldots,x_{n})\Phi(x_{1})\ldots\Phi(x_{n}) \ d^{4}x_{1}\ldots d^{4}x_{n}\\
\Gamma^{(n)} & (x_{1},\ldots,x_{n})\equiv\frac{1}{\sqrt{-g(x_{1})}\ldots\sqrt{-g(x_{n})}}\frac{\delta^{n}\Gamma[\Phi]}{\delta\Phi(x_{1})\ldots\delta\Phi(x_{n})}\Big|_{\Phi(x)=0},
\end{split}
\end{equation} 
where $\Gamma^{(n)}(x_{1},\ldots,x_{n})$ is nothing but the sum of all one particle irreducible diagrams with $n$ external lines.

From the definition of generating functional, it follows that effective can also be expressed as
\begin{equation}
\Gamma[\Phi]=-\int\sqrt{-g(x)}d^{4}x\Big[\mathbf{Z}[\Phi]\frac{1}{2}g^{\mu\nu}\partial_{\mu}\Phi(x)\partial_{\nu}\Phi(x)+\mathcal{V}_{\text{eff}}[\Phi]\Big],
\end{equation}
where $\mathbf{Z}[\Phi]$ is the wavefunction renormalization constant and $\mathcal{V}_{\text{eff}}[\Phi]$ is known as the effective potential which is sum of all 1PI diagrams with zero external momenta. This follows from the expression in (\ref{series effective action})
\begin{equation}
\begin{split}
\Gamma^{(n)}(x_{1},\ldots,x_{n}) & =\int\frac{d^{4}p_{1}}{(2\pi)^{4}}\ldots\frac{d^{4}p_{n}}{(2\pi)^{4}} e^{i(p_{1}.x_{1}+\ldots+p_{n}.x_{n})}\Gamma^{(n)}(p_{1},\ldots,p_{n})(2\pi)^{4}\delta^{(4)}(p_{1}+\ldots+p_{n})\\
\Gamma^{(n)}(p_{1},\ldots,p_{n}) & =\Gamma^{(n)}(0,\ldots,0)+\text{derivative expansion}\\
\implies\Gamma[\Phi] & =\int\sqrt{-g(x)}d^{4}x\sum_{n=0}^{\infty}\frac{1}{n!}\int\sqrt{-g(x_{1})}\ldots\sqrt{-g(x_{n})}d^{4}x_{1}\ldots d^{4}x_{n} \ \Phi(x_{1})\ldots\Phi(x_{n})\\
 & \times\int\frac{d^{4}p_{1}}{(2\pi)^{4}}\ldots\frac{d^{4}p_{n}}{(2\pi)^{4}} e^{i(p_{1}.x_{1}+\ldots+p_{n}.x_{n})}e^{-i(p_{1}+\ldots+p_{n}).x}(\Gamma^{(n)}(0,\ldots,0)+\ldots)\\
 & =\int\sqrt{-g(x)}d^{4}x\sum_{n}\frac{1}{n!}[\Gamma^{(n)}(0,\ldots,0)(\Phi(x))^{n}+\ldots],
\end{split}
\end{equation}
where $\mathcal{V}_{\text{eff}}[\Phi]=\int\sqrt{-g(x)}d^{4}x\sum_{n}\frac{1}{n!}\Gamma^{(n)}(0,\ldots,0)(\Phi(x))^{n}$ (see the Fourier expansion techniques introduced in \cite{mandal2019local}).

\subsection{Loop expansion of effective action}
From the definitions (\ref{generating0}), (\ref{generating2}) and (\ref{effective action}), following relation can be shown easily
\begin{equation}\label{effective action2}
e^{i(\Gamma[\Phi]+\int\sqrt{-g(x)}d^{4}x \ J(x)\Phi(x))}=\int\mathcal{D}\phi \ e^{iS[\phi]-i\int\sqrt{-g(x)}d^{4}x \ \frac{\delta\Gamma[\Phi]}{\delta\Phi(x)}\phi(x)},
\end{equation}
and doing a shift in integration variable $\phi\rightarrow\phi+\Phi$, above relation can be expressed as
\begin{equation}
e^{i\Gamma[\Phi]}=\int\mathcal{D}\phi \ e^{i\left(S[\phi+\Phi]-\int\sqrt{-g(x)}d^{4}x \ \frac{\delta\Gamma[\Phi]}{\delta\Phi(x)}\phi(x)\right)}.
\end{equation}  
$S[\phi+\Phi]$ can be expressed as a Taylor series expansion which is as follows
\begin{equation}
\begin{split}
S[\phi+\Phi] & =S[\Phi]+\sum_{n=1}^{\infty}\frac{1}{n!}\int\sqrt{-g(x_{1})}d^{4}x_{1}\ldots\sqrt{-g(x_{n})}d^{4}x_{n}S^{(n)}(x_{1},\ldots,x_{n}|\Phi)\phi(x_{1})\ldots\phi(x_{n})\\
 & \equiv S[\Phi]+\sum_{n=1}^{\infty}\frac{1}{n!}S_{n}[\Phi]\phi^{n}\\
S^{(n)} & (x_{1},\ldots,x_{n}|\Phi)=\frac{\delta^{n}S[\Phi]}{\delta\Phi(x_{1})\ldots\delta\Phi(x_{n})}, 
\end{split}
\end{equation}
and plugging above expression in (\ref{effective action2}), we obtain
\begin{equation}\label{effective action3}
e^{i(\Gamma[\Phi]-S[\Phi])}=\int\mathcal{D}\phi \ e^{i\left(S_{2}[\Phi]\phi^{2}+\sum_{n=3}^{\infty}\frac{1}{n!}S_{n}\phi^{n}-\phi(\Gamma_{1}[\Phi]-S_{1}[\Phi])\right)}
\end{equation}
where $\Gamma_{1}[\Phi]=\frac{1}{\sqrt{-g(x)}}\frac{\delta\Gamma[\Phi]}{\delta\Phi(x)}$ and $\phi\Gamma^{(1)}[\Phi]\equiv\int d^{4}x\frac{\delta\Gamma[\Phi]}{\delta\Phi(x)}\phi(x)$.

Defining $\bar{\Gamma}[\Phi]\equiv\Gamma[\Phi]-S[\Phi]$ and putting back $\hbar$ explicitly exponents in (\ref{effective action3}), the expression becomes
\begin{equation}
e^{\frac{i}{\hbar}\bar{\Gamma}[\Phi]}=\int\mathcal{D}\phi \ e^{i\left(S_{2}[\Phi]\phi^{2}+\sum_{n=3}^{\infty}\frac{\hbar^{\frac{n}{2}-1}}{n!}S_{n}[\Phi]\phi^{n}-\hbar^{-\frac{1}{2}}\phi\bar{\Gamma}_{1}[\Phi]\right)}.
\end{equation}
Considering an expansion of the form $\bar{\Gamma}[\Phi]=\sum_{n=1}^{\infty}\hbar^{n}\bar{\Gamma}^{(n)}[\Phi]$, we can find each coefficient in this expansion by doing a loop expansion i.e. expansion in $\hbar$ of following relation
\begin{equation}
e^{i\sum_{n=1}^{\infty}\hbar^{n-1}\bar{\Gamma}^{(n)}[\Phi]}=\int\mathcal{D}\phi \ e^{i\left(S_{2}[\Phi]\phi^{2}+\sum_{n=3}^{\infty}\frac{\hbar^{\frac{n}{2}-1}}{n!}S_{n}[\Phi]\phi^{n}-\sum_{n=1}^{\infty}\hbar^{n-\frac{1}{2}}\phi\bar{\Gamma}_{1}^{(n)}[\Phi]\right)}.
\end{equation}
In the leading order $\mathcal{O}(\hbar)$, we obtain
\begin{equation}\label{effective potential1}
\begin{split}
\bar{\Gamma}^{(1)}[\Phi] & =\frac{i}{2}\log\text{det}[S_{2}[\Phi]]\\
\implies\Gamma[\Phi] & =S[\Phi]+\hbar\frac{i}{2}\log\text{det}[S_{2}[\Phi]]+\mathcal{O}(\hbar^{2})\\
 & =S[\Phi]+\hbar\frac{i}{2}\text{Tr}\log[S_{2}[\Phi]]+\mathcal{O}(\hbar^{2})\\
\implies\mathcal{V}_{\text{eff}}[\Phi] & =V_{0}[\Phi]-\hbar\frac{i}{2}\text{Tr}\log[S_{2}[\Phi]]+\mathcal{O}(\hbar^{2}). 
\end{split}
\end{equation}

\subsection{Coleman-Weinberg potential for massless $\phi^{4}$ field theory}
Action for a massless scalar field theory with $\phi^{4}$-interaction is given by
\begin{equation}
S[\phi]=-\int\sqrt{-g}d^{4}x\Big[\frac{1}{2}g^{\mu\nu}\partial_{\mu}\phi\partial_{\nu}\phi+\frac{\lambda}{4!}\phi^{4}+\frac{A}{2}g^{\mu\nu}\partial_{\mu}\phi\partial_{\nu}\phi+\frac{B}{2}\phi^{2}+\frac{C}{4!}\phi^{4}\Big],
\end{equation}
where $A, \ B, \ C$ are the counterterms, required in order to renormalize the field theory. From (\ref{effective potential1}), we can express the effective potential as
\begin{equation}
\mathcal{V}_{\text{eff}}[\Phi]=\frac{\lambda}{4!}\Phi^{4}-\frac{B}{2}\Phi^{2}+\frac{C}{4!}\Phi^{4}+\frac{i}{2}\text{Tr}\log[\mathcal{G}[\Phi]],
\end{equation} 
where
\begin{equation}
\begin{split}
\frac{i}{2}\text{Tr}\log[\mathcal{G}[\Phi]] & =\frac{i}{2}\text{Tr}\log[\mathcal{G}_{0}[\Phi]+\mathcal{G}_{2}[\Phi]+\mathcal{G}_{3}[\Phi]+\mathcal{G}_{4}[\Phi]+\ldots]\\
 & =-\frac{i}{2}\text{Tr}\log[\mathcal{G}_{0}[\Phi]]+\frac{i}{2}\text{Tr}\log[1+\tilde{\mathcal{G}}_{2}[\Phi]+\tilde{\mathcal{G}}_{3}[\Phi]+\tilde{\mathcal{G}}_{4}[\Phi]+\ldots]\\
 & =-\frac{i}{2}\int\frac{d^{4}k}{(2\pi)^{4}}\log[\mathcal{G}_{0}^{-1}[k,\Phi]]+\frac{i}{2}\int\frac{d^{4}k}{(2\pi)^{4}}\Big[\tilde{\mathcal{G}}_{2}[k,\Phi]+\tilde{\mathcal{G}}_{4}[k,\Phi]-\frac{1}{2}\tilde{\mathcal{G}}_{2}^{2}[k,\Phi]\Big].
\end{split}
\end{equation}
Using the Wick rotation, we express first two terms in the series expansion since the third and fourth terms are independent of $\Phi$. Using dimensional regularization, first term can be expressed as follows
\begin{equation}
\begin{split}
-\frac{i}{2}\int\frac{d^{4}k}{(2\pi)^{4}} & \log[\mathcal{G}_{0}^{-1}[k,\Phi]]=\frac{1}{2}\int\frac{d^{4}k}{(2\pi)^{4}}\log[k^{2}+\frac{\lambda}{2}\Phi^{2}]\\
 & =\frac{1}{64\pi^{2}}\Gamma\left(-1+\frac{\epsilon}{2}\right)\left(\frac{\lambda}{2}\Phi^{2}\right)^{2-\frac{\epsilon}{2}}\\
 & =\frac{\lambda^{2}}{256\pi^{2}}\Phi^{4}\left(-\frac{2}{\epsilon}+\psi(2)+\mathcal{O}(\epsilon)\right)\left(1-\frac{\epsilon}{2}\log\left(\frac{\lambda\Phi^{2}}{2\Lambda^{2}}\right)+\mathcal{O}(\epsilon^{2})\right)\\
 & =-\frac{\lambda^{2}}{128\pi^{2}\epsilon}\Phi^{4}+\frac{\lambda^{2}}{256\pi^{2}}\Phi^{4}\log\left(\frac{\lambda\Phi^{2}}{2\Lambda^{2}}\right)+\frac{\lambda^{2}}{256\pi^{2}}\Phi^{4}\psi(2),
\end{split}
\end{equation}
whereas the second term (leading order curvature correction) becomes
\begin{equation}
\begin{split}
\frac{i}{2}\int\frac{d^{4}k}{(2\pi)^{4}} & \tilde{\mathcal{G}}_{2}[k,\Phi]=-\frac{1}{6}\mathcal{R}\int\frac{d^{4}k}{(2\pi)^{4}}\frac{1}{k^{2}+\frac{\lambda}{2}\Phi^{2}}=-\frac{1}{192\pi^{2}}\mathcal{R}\Gamma\left(-1+\frac{\epsilon}{2}\right)\left(\frac{\lambda}{2}\Phi^{2}\right)^{1-\frac{\epsilon}{2}}\\
 & =-\frac{\lambda}{384\pi^{2}}\mathcal{R}\Phi^{2}\left(-\frac{2}{\epsilon}+\psi(2)+\mathcal{O}(\epsilon)\right)\left(1-\frac{\epsilon}{2}\log\left(\frac{\lambda\Phi^{2}}{2\Lambda^{2}}\right)+\mathcal{O}(\epsilon^{2})\right)\\
 & =\frac{\lambda}{192\pi^{2}\epsilon}\mathcal{R}\Phi^{2}-\frac{\lambda}{192\pi^{2}}\mathcal{R}\Phi^{2}\log\left(\frac{\lambda\Phi^{2}}{2\Lambda^{2}}\right)-\frac{\lambda}{384\pi^{2}}\mathcal{R}\Phi^{2}\psi(2).
\end{split}
\end{equation}
Thus, the effective potential becomes the following after choosing the appropriate counterterms 
(maintaining the renormalizability conditions at arbitrary energy scale $M$)
\begin{equation}\label{modified coleman-weinberg}
\begin{split}
\mathcal{V}_{\text{eff}}[\Phi] & = \frac{\lambda}{4!}\Phi^{4}+\frac{\lambda^{2}}{256\pi^{2}}\Phi^{4}\Bigg[\log\left(\frac{\Phi^{2}}{M^{2}}\right) - \frac{25}{6}\Bigg]\\
 & + \frac{\lambda \mathcal{R}}{48\pi^{2}}\Phi^{2}\Bigg[1 - \frac{1}{4}\log\left(\frac{\Phi^{2}}{M^{2}}
 \right)\Bigg] - \frac{\lambda \mathcal{R}}{1152\pi^{2} M^{2}}\Phi^{4}.
\end{split}
\end{equation}
%
%
%
%
This clearly shows the effect of curved spacetime on the Coleman-Weinberg potential. Further, 
the presence of Ricci-scalar $\mathcal{R}$ dependent correction to Coleman-Weinberg potential \cite{coleman1973radiative} in (\ref{modified coleman-weinberg}) also shifts the value of 
non-zero vacuum expectation value of the scalar field $\Phi_{c}$. It is also important to note 
here that in the Minkowski spacetime, the above-mentioned correction to Coleman-Weinberg potential
vanishes and the potential reduces to the standard result in the literature.

\section{Discussion}
Extending the RNC formalism of QFT in curved spacetime \cite{mandal2019local} to thermal field theories in curved spacetime using the imaginary time formulation and assuming local thermal equilibrium, we are able to bring out the curvature-induced corrections to physical observables in these field theories. Some of the physical observables are computed up to few leading terms in order to show curvature-dependent corrections explicitly. Apart from the computation of physical observables, we also show the consequence of these curvature-corrections to the BEC phenomenon and derive a constraint from the positivity of the critical temperature of BEC. Further, it is also shown that even at zero-temperature conserved 4-current associated with global $U(1)$-symmetry is non-zero which has potential implications in astrophysical compact objects since it determines the electric and magnetic field distributions from Maxwell's equations. Since SSB is an important phenomenon that explains the origin of superfluidity, BEC, and other novel phases of matter, we also discuss the curvature-induced corrections to Coleman-Weinberg potential in massless $\phi^{4}$-theory.

Lastly, we want to mention the domain of validity of RNC technique used in thermal field theories coupled to curved spacetimes. Since RNC covers a patch in the neighborhood of a given point, which will be taken as the origin of this coordinate system, this coordinate system is well-defined as long as the geodesics do not intersect in that patch. These coordinates are constructed using geodesics emanating from $p$ (an arbitrary point), identified with straight lines in the tangent space $T_p M$ ($M$ being spacetime manifold). Validity requires that the exponential map $\exp_p : T_p M \to M$ is smooth and invertible in the chosen region. RNC are valid only in a \emph{normal neighborhood} of $p$, i.e.\ a region where: (i) every point is connected to $p$ by a unique geodesic, and (ii) The exponential map is a diffeomorphism. The above domain of validity breaks down at \emph{conjugate points}, where distinct geodesics from $p$ intersect. Beyond these points, the exponential map ceases to be injective. Therefore, the maximal domain of RNC ends at the \emph{cut locus} of $p$, the set of points where minimizing geodesics from $p$ are no longer unique. A manifold need not be geodesically complete to allow RNC at a point; only a sufficiently small normal neighborhood is required. If the point $p$ is too close to a curvature singularity or a boundary, the normal neighborhood may shrink to zero size, limiting RNC validity. Hence, RNC is valid over a region where tidal effects (from curvature) are small enough that quadratic curvature corrections remain controlled. The above mentioned points mostly hold for spacetimes with relatively small curvature. 

\section{Acknowledgement}
SM is supported by IISER Tirupati through a post-doctoral fellowship.

\bibliographystyle{unsrt}
\bibliography{draft}

\end{document}